# Annotations, Collaborative Tagging, and Searching Mathematics in E-Learning

Iyad Abu Doush
Department of Computer Sciences
Yarmouk University
Irbid, Jordan

Faisal Alkhateeb
Department of Computer Sciences
Yarmouk University
Irbid, Jordan

Eslam Al Maghayreh
Department of Computer Sciences
Yarmouk University
Irbid, Jordan

Izzat Alsmadi
Department of Computer Information Systems
Yarmouk University
Irbid, Jordan

Samer Samarah
Department of Computer Information Systems
Yarmouk University
Irbid, Jordan

*Abstract*—This paper presents a new framework for adding semantics into e-learning system. The proposed approach relies on two principles. The first principle is the automatic addition of semantic information when creating the mathematical contents. The second principle is the collaborative tagging and annotation of the e-learning contents and the use of an ontology to categorize the e-learning contents. The proposed system encodes the mathematical contents using presentation MathML with RDFa annotations. The system allows students to highlight and annotate specific parts of the e-learning contents. The objective is to add meaning into the e-learning contents, to add relationships between contents, and to create a framework to facilitate searching the contents. This semantic information can be used to answer semantic queries (e.g., SPARQL) to retrieve information request of a user. This work is implemented as an embedded code into Moodle e-learning system.

I. INTRODUCTION

In recent years, we have observed the fast evolution of the Internet. The advent of the Internet has significantly enhanced availability of technical content, by making millions of documents available on-line. The introduction of sophisticated Learning Management Systems (LMSs), such as Blackboard and Moodle, has widely increased the opportunities for a geographically diverse population of students to access educational programs.

Recent statistics (Whitehead, 2009) indicate that more than 3.2 million people are involved in some form of online education. Modern LMSs are designed to provide sophisticated organization of the various components of a course.

The literature provides several proposals aimed at adding semantic meaning/markups to different parts of the online course (e.g., (Bateman, 2007; Bateman, Brooks, Mccalla, & Brusilovsky, 2007; Stojanovic, Staab, & Studer, 2001)). We strongly believe that this can have a profound impact in facilitating the use of LMSs.

Partitioning the course page into logically related components will allow the student to select specific parts of the learning contents (Enagandula, Juthani, Ramakrishnan, Rawal, & Vidyasagar, 2005). This would minimize the learning time and direct the user to focus on the parts that s/he is interested in.

Following a research trend that has developed in the last years, we explore the introduction of technologies drawn from the field of the semantic web to aid in the process of semantic searching of LMSs. We rely on three principles: ontologies, RDFa annotations, and folksonomies.

Ontology is one of the main elements of the semantic web. An ontology is used to formalize the concepts in a domain and to describe their mutual relationships (Kim, Kim, & Park, 2005). Ontologies can be used to build adaptable e-learning systems: using ontologies, human or software agents can automatically understand the meaning of the e-learning content and process it appropriately (Kim et al., 2005). For example, ontologies enable semantic-based search of e-learning content (Chang, Ham, Moon, Choi, & Cha, 2007).

The process of collaborative tagging produces a set of tags that can be used to describe a resource (Al-Khalifa & Davis, 2006). This tagging process, made popular by sites like del.icio.us, can be employed to generate grassroot ontologies, commonly called folksonomies. Folksonomies add semantics to resources through a social markup process, enabling the use of tag clouds to describe entities and determine relevancy. The use of folksonomies allows the addition of semantics without the aid of individual manual indexers or automated keyword generators (Al-Khalifa & Davis, 2006).

The tagging process does not create a strict taxonomy for objects, but it enables the user to employ his own keywords to categorize objects. Gruber (Gruber, 2005) mentioned that, in collaborative tagging, there is no explicit links between entities, and no standard form is used for presenting the data. This limitation requires introducing a formal ontology along with folksonomies to formalize tags.

Resource Description Framework (RDF) data can be embedded inside XHTML as RDFa ("RDFa Primer: Bridging the Human and Data Webs, http://www.w3.org/TR/xhtml-rdfa-primer/." 2010). The RDFa annotations are used for making parts of the web page foldable into a more detailed information (i.e., according to the vocabulary and the relations of the used RDF). Standard extractors for RDFa can be used to retrieve the annotations in the web page (e.g., ("RDFa Distiller and Parser, http://www.w3.org/2007/08/pyRdfa/." 2010; "rdfquery, RDF processing in your browser, http://code.google.com/p/rdfquery/." 2010)).

Mathematical contents represent a particular challenge for searching the contents in e-learning - e.g., formulas, mathematical symbols, and abbreviated function names. The W3C recommendation for encoding mathematics on the web is called MathML("Mathematical Markup Language, http://www.w3.org/Math/." 2010). There are two types of MathML: presentation MathML and contents MathML. Presentation MathML describes the visual appearance of the mathematical expression by using 2-dimensional layout and formatting of the mathematical expression. On the other hand, content MathML encodes the meaning or the mathematical semantic of the expression.

The encoding of mathematical expressions using presentation MathML can help in capturing the conceptual structure of the mathematical expression. This can help in having a common notation which can be used for the underlying search of the mathematical contents.

Adding semantics to the e-learning contents on the web can provide several benefits to the users:

1. Provide a more accessible contents for the blind and visually impaired individuals, as the contents can be read by screen readers.

2. Easier searching for technical and educational materials.

3. The contents can be explained (e.g., using other students annotations or from relations between other contents).

4. Help people with learning disabilities in navigating the e-learning contents (e.g., providing information about a concept in the navigated e-learning as a tooltip).

In this paper we propose a novel framework for adding semantic information to the e-learning contents. The e-learning contents can be highlighted, and annotations and tags can be added by the student. A defined ontology is used to categorize user annotations and tags. Another part of the framework is the automatic insertion of semantic information to the mathematical equation. The mathematical contents in e-learning are annotated by semantic information using RDFa.

## II. BACKGROUND

### A. Searching Mathematical Contents

A review of related work shows that developing searchable mathematical expressions imposes a number of requirements. An extensive literature exists, dealing with various aspects of this problem. Some of these relevant studies are discussed next.

Munavalli and Miner (Munavalli & Miner, 2006) introduce a math aware search engine which search mathematical contents. The system analyzes MathML mathematical expressions into text math fragments. In this system the user enters the math query using graphical equation editor.

According to Youssef (Youssef, 2006) the mathematical search purpose is: 1) allowing the user to perform fine grained search on mathematical data 2) Allow users to enter the math query naturally and easily using the symbols and notations applied by mathematicians and scientists.

In another work Guidi and Schena (Guidi & Schena, 2003) introduce a math query language for RDF metadata repository called MathQL. Asperti et al. (Asperti, Padovani, Coen, & Schena, 2001) presented HELM, a framework that uses XML technology for building structures contents in a logical manner. The purpose is to use the system as a library for indexing and retrieving mathematical documents.

Altamimi and Youssef (Altamimi & Youssef, 2008) presented a math query language that enable users to express their information needs intuitively yet precisely. The new math query language offers an alternative way to describe mathematical expressions that are more consistent and less ambiguous than conventional mathematical notations. In addition, the language goes beyond the Boolean and proximity query syntax found in standard text search systems. It defines a powerful set of wildcards that are deemed important for math search. These wildcards provide precise structural search and multi-levels of abstractions.

Hijikata et al. (Hijikata, Hashimoto, & Nishida, 2009) presented a search engine for MathML objects using the structure of mathematical formulas. The system makes the inverted indices by using the Document Object Model (DOM) structure of the MathML object. It also proposes three types of indexes:

One type is constructed from some paths of the DOM structure and expressed in XPath.

The second type is constructed by encoding the nodes in the same level in DOM structure.

The third type is a hybrid method from the other two types.

### B. Semantic Web and E-learning

An approach based on a student model and an ontology in order to personalize an eLearning system is proposed in (Gomes, Antunes, Rodrigues, Santos, & Barbeira, 2006; Pah, Stoica, Cacovean, & Popa, 2008). The ontology is used to map the student knowledge to course concepts allowing a better access to her/his progress and to adapt contents and navigation structure to a particular student.

Soylu et al. (Soylu, Kuru, Wild, & Mdritscher, 2008) proposed a framework for harvesting learning objects from web-based content. The framework is based on a lightweight application profile and a microformat for learning objects using well-known learning object metadata standards in order to address mainly interoperability and reusability of learning content. They also describe a web service to extract learning objects from different web pages, and provide an SQI target as a retrieval facility using SPARQL and XSL transformation.

Henze et al. (Henze, Dolog, & Nejdl, 2004) proposed an approach for personalization of e-Learning systems by using semantic web and shows how the semantic web resource description formats can be utilized for automatic generation of hypertext structures from distributed RDF annotations. Several ontologies have been utilized corresponding to the components of an adaptive hypermedia system: a domain ontology (describing the document space, the relations of documents, and concepts covered in the domain of this document space), a user ontology (describing learner characteristics), and an observation ontology (modeling different possible interactions of a user with the hypertext).

In another work Stojanovic et al. (Stojanovic et al., 2001) presented an approach for implementing the eLearning scenario using Semantic Web technologies. The goal of the proposed framework is to provide a flexible and personalized access to these learning Materials, the proposed eLearning scenario exploit ontologies in three ways: for describing the semantics (content) of the learning materials (this is the domain dependent ontology), for defining the learning context of the learning material and for structuring the learning materials in the learning courses.

*C. Collaborative Tagging and E-learning*

In another work Stojanovic et al. (Stojanovic et al., 2001) presented an approach for implementing the eLearning scenario using Semantic Web technologies. The goal of the proposed framework is to provide a flexible and personalized access to these learning Materials, the proposed eLearning scenario exploit ontologies in three ways: for describing the semantics (content) of the learning materials (this is the domain dependent ontology), for defining the learning context of the learning material and for structuring the learning materials in the learning courses.

Collaborative tagging (also known as folksonomy, social classification, social indexing, and social tagging) is a system for collaboratively creating and managing tags to annotate and categorize contents. The term folksonomy was first used by Thomas Vander Wal in a discussion on an information architecture mailing list. It is a combination of "folk" and "taxonomy". (Folks => done by people, Taxonomy => classification of items into groups) (Peters & Becker, 2009).

One of the main objectives of collaborative tagging is to make a collection of information increasingly easy to search, discover, and navigate over time. Thus, collaborative tagging has a main role in developing semantic web and information retrieval systems.

Collaborative tagging became popular on the Web around 2004 as part of social software applications such as social bookmarking and photograph annotation. Tagging, which is one of the defining characteristics of Web 2.0 services, allows users to collectively classify and find information. Some websites include tag clouds as a way to visualize tags in a folksonomy (Mathes, 2004; Peters & Becker, 2009).

Bateman et el. (Bateman, 2007) outline their experience with applying collaborative tagging in e-learning systems to supplement more traditional metadata gathering approaches. They state that metadata is best created if it focuses on a particular goal, is contextualized to a particular user, and is created in an ambient manner by observing the actions and interactions of students in learning environments. Consequently, collaborative tagging seems to be a leading method by which we can collect learner-centric metadata. Using tags enables useful resource organization and browsing techniques, and the viewing of tags used on a webpage can give a learner some idea of its importance and its content, it may help a learner in finding the exact point of interest within the page.

According to (Bateman, 2007) collaborative tagging systems have potential to be a good fit with e-learning systems, because of the following:

1. Learning managements systems currently lack sufficient support for self organization of learning content.

2. Collaborative tagging has potential to further enrich peer interactions and peer awareness centered around learning content.

3. Tagging, by its nature is a reflective practice, which can give students an opportunity to summarize new ideas, while receiving peer support (through viewing other learners' tags; tag suggestions).

4. The information provided by tags provides insight on learner's comprehension and activity, which is useful for both educators and administrators.

Bateman et el. (Bateman, Brooks, & Mccalla, 2006) have suggested merging the use of ontologies with collaborative tagging through a new approach and they have called it CommonFolks. This approach should help to reduce the effort required in the production of useful metadata, while maintaining the expressiveness inherent in lightweight ontologies, thus opening the door to a better quality of metadata and authoring by those not traditionally involved in metadata creation.

In (Bateman, 2007), Bateman states that the interest in investigating the use of tagging in e-learning revolves around several points of interest. Some of these points can be summarized as follows:

- Organization and Annotation: learning management systems (LMSs) do not properly support self organization and annotation of learning content. However, students usually use a number of organization and annotation techniques. These include writing notes, creating marginalia in books, highlighting text, and bookmarking pages. How can we make these traditional capabilities available to students using digital learning materials? Are there methodologies more suitable for such tasks in an online environment?

- In this regard, tagging and note-taking can be linked together, since tags represent an aspect to be used in the tagger's recall process. Tagging provides a straightforward method for self-organization and most tagging interfaces provide fields for longer sentence based notes to be taken (Bateman, 2007).

- Metadata Collection: according to the learning object paradigm, online educational content can be collected, aggregated, and packaged for delivery to learners. However, the major difficulty in accomplishing this vision is the lack of meaningful metadata describing learning objects. Collaborative tags can be considered as a form of metadata, which could supplement the needs for detailed learning object descriptions (Bateman, 2007).

- Knowledge Gain: tagging gives students the chance to summarize new ideas while receiving peer support. Tagging can be considered as an action of reflection, where the tagger can summarize a series of thoughts into one or more tags, each of which stands on its own to describe some aspect of the resources based on the tagger's experiences and beliefs (Bateman, 2007).

- Pedagogical Reflection: in e-learning there is a lack of direct interaction with learners that inform instructors about the understanding of new concepts by them. Collaborative tags, created by learners can help instructors in predicting their students' progress e.g. tags that are out of context could represent a misconception (Bateman, 2007).

### III. METHODOLOGY

We have developed an ontology and infrastructure for the annotation of the course content in Moodle. The proposed ontology consists of activities and entities involved in the e-learning process (e.g., lesson, quiz, and student). The ontology defines the relationships between different classes in the e-learning ontology (e.g., an assignment is solved by a student). The ontology can be collaboratively extended by the users of the system by introducing new classes that are associated with other pre-defined classes in the ontology.

The ontology is structured in two parts: an upper ontology and a bottom ontology. Intuitively, the upper ontology is used to describe the overall structure of a generic course in Moodle. The upper ontology describes the components of a course that are common across all courses; as such, this is a static ontology. The bottom ontology describes the components that are course-specific and domain-specific. It is an inherently dynamic ontology.

The fundamental idea is to get the help from students in the class to introduce a taxonomy for a specific online course. Students participating in the same class are encouraged to annotate the course contents and share such annotations with other students. The benefit from integrating the two approaches (i.e., bottom-up and top-down) is to create a comprehensive semantic structure of a course, which can be used by students to understand the content of the course and facilitate the semantic searching of the contents.

For example, a student who reads the online course material can add annotations to explain specific parts of the lecture notes (e.g., describe a figure or a mathematical formula) in the process of studying the material. These annotations can be used by other students; the tagging of a figure can, for example, provide a textual documentation to look for the figure in the online course.

We use the current web page in Moodle (e.g. quiz, lesson...) to select the ontology class for the highlight. We will suggest the category (class) for the user tags using synonyms and other user's tags. WordNet synonyms are used to infer concepts related to the students' tags. Synonyms are used to group tags as instances of the same ontology concept.

An equation editor is used to create mathematical equations encoded in presentation MathML. The mathematical contents are then annotated using RDFa with other set of semantic information (i.e., link, value, description, and category) which will be explained next in the paper. The main goal is to provide the user with different search schemes (i.e., keywords, equation structure, equation category, and equation uses).

*A. Annotation System Use Scenario*

To understand how the system operates, let us consider a student, who wants to search an online course in Moodle to review lecture notes for a math course. (e.g., a mathematical formula is annotated with a link that explains it, or a paragraph in lecture notes is tagged as a potential quiz question).

The user can use the category of the tag to find its relation with other course materials (e.g., a specific part of the lecture notes is tagged as an explanation for a mathematical formula in the notes the student is browsing). When the student encounters graphical content in the page he could look for the annotations associated with it to get explanation about its contents and how it relates to the content of the lecture notes (e.g., a graph for mathematical equation). The use of the taxonomy provided by the ontology can be seen also when the student wants to access specific parts of the page (e.g., geometry → area of a triangle).

*B. The Top Ontology*

We built different classes in the top ontology (Figure1) on the basis of various components of Moodle (e.g., quiz, homework). The Activity class is the super class of all the activities (Lesson, Glossary, Assignment, Wiki, Forum, and Resource) the student can perform in the e-learning system. Smaller components within the e-learning activity are defined (e.g., Post in a forum, Question in a quiz) to give a category for smaller components of the annotated information.

The interactions between the different components in the LMS are presented as relationships between different concepts

(e.g., Quiz ConsistOf Questions, Assignment isSolvedBy Student). The ontology concepts and the relationships between them are used to describe the structure of a course and as main concepts to be refined by the students with their bottom-ontology.

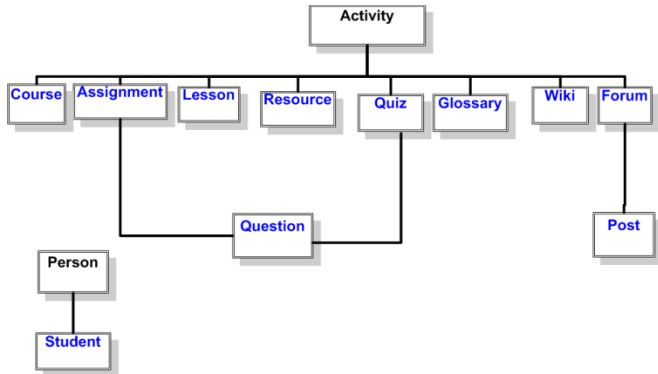

Figure1- Graph of a Fragment of the Upper Ontology

*C. The Bottom Ontology*

The user can tag a course component (e.g., a text fragment, an image, a formula) in the online course. The predefined set of upper ontology concepts can be used as the categories for the current annotated resource, or the user can define new concepts that are used to refine and expand the upper-ontology. Each tag introduced by the student is presented as an instance of the concept selected or defined by the student.

For example, a student can define a new concept (e.g., trigonometry formula) to annotate an equation in a math course lesson. The annotated equation will become an instance of this new concept. The user can add a tag or a set of tags (e.g., math, formula) to the annotated equations which are linked to the selected class via the has-tag relationship. As another example, the student can annotate a post in a forum as an explanation for a specific lesson; in this case, the annotated post will be treated as an instance of the Post class, and the connection between the Post and Lesson classes will be identified using the relationship defined between the two classes.

*D. Generating the Bottom Ontology*

Collaborative tagging among different students of an online course can help in extending the upper-ontology and create new connections between different learning contents. Users are allowed to generate tag clouds, by freely tagging the content encountered in Moodle.

The tag cloud for a component of a Moodle class will provide domain specific classification of the elements of the course as well as the level of ``agreement'' across students concerning the description of the course entities. The tags frequency can be used as an agreement measure (based on the algorithm in (Heymann & Garcia-Molina, 2006)). The most agreed upon tags will flow into the bottom ontology, as new concepts. Using the proposed system (Figure2) the user can highlight a resource in the online course (e.g., text fragment) and then the tool interface can be used to select then category (concept) of the annotated resource or to create a new concept.

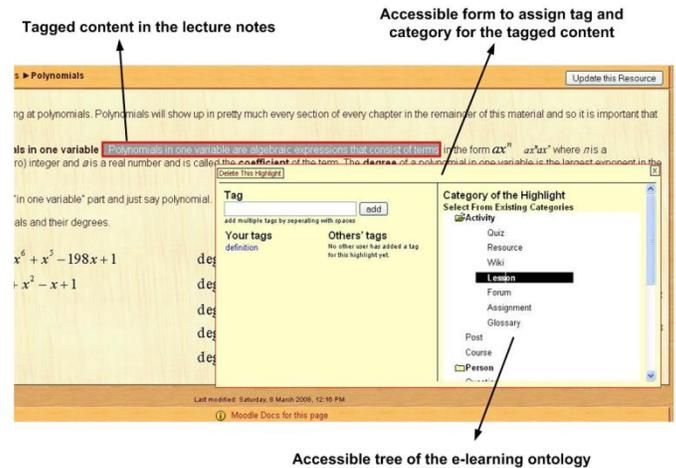

Figure2- A Screenshot of the Proposed System

The construction of the tags and the expansion of the ontology is user-driven. The user can select the concept for the highlighted content and assign a tag for the content. The user can navigate the tags (i.e., perform keyword search) in order to retrieve the set of items associated with that tag.

*E. A Mathematical Search Use Scenario*

The e-learning content creator uses an equation editor inside the learning management system to generate mathematical contents. The content creator adds other information (i.e., description) to the mathematical contents. Other information is also added automatically to the mathematical contents (i.e., category, value, source, and link). The created contents are then annotated with this information using RDFa.

A student who is interested in searching for integral examples in lessons only can then search the proposed system by using the keyword "integral" and tick the checkbox lessons. The results can be obtained using extracted RDFa annotations. The student can also specify a specific type of integrals to search for by using the equation editor to enter a specific integral and search for it in specific place in the e-learning contents (e.g., previous quizzes). The system returns a URL that points the user exactly where the searched mathematical formula is located in the e-learning content.

*F. Mathematical Contents with Semantic Information*

The equation editor can be used by the mathematical contents generator in e-learning to create a mathematical equations encoded in presentation MathML. The presentation MathML can be visually rendered by the machines but they are not understandable by them (Asperti et al., 2001). Presentation MathML does not provide adequate semantic information (Kohlhase & Sucan, 2006).

The encoding of mathematical expressions using common RDFa annotations can help in capturing the conceptual structure and remove any ambiguity and inconsistency related to the use of presentation MathML in encoding the mathematical expression. This can help in having a common notation which can be used for the underlying search of the mathematical contents.

The goal of this part of the system is to provide the student with a mathematical content with semantic encoding. This can lead to a more accurate search of the mathematical contents in e-learning. Using this scheme the system will search for the agreed vocabulary used in RDFa.

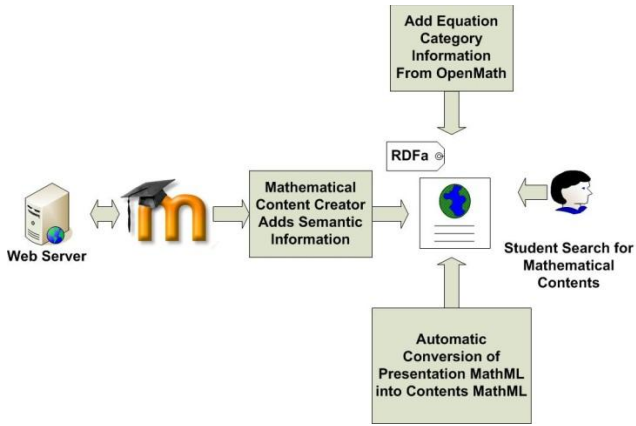

Figure3- The System for Adding Semantic Information to the Mathematical Contents.

- The proposed methodology consists of the following major phases (see Figure3):

- Mathematical content creator in e-learning uses an equation editor to build a mathematical equation. The equation is encoded using presentation MathML.

- The contents creator adds other parts (e.g., description) to the content as RDFa annotations.

- In order to allow the user to do a fine grain search on the page level the presentation MathML is annotated using RDFa.

- The RDFa extractors will be used next to extract the information about the mathematical expressions along with the URI of the mathematical expression.

- The extracted RDFa annotations can be compared then with the user query to retrieve mathematical contents with closer similarity.

We will have a place where the mathematical vocabulary (i.e., classes and properties) located, and use this vocabulary to annotate the content MathML. The content MathML from the previous example can be annotated as follows:

```
<div xmlns:m="http://example.com/math/vocab#">
  <div about="/math101/lesson1/powerEquation">
    <math>
      <apply >
        <power property="m:binary-operator"/>
        <apply>
          <plus  property="m:unary-operator"/>
          <ci property="m:identifier">x</ci>
          <cn property="m:number">3</cn>
        </apply>
        <cn property="m:number">2</cn>
      </apply>
    </math>
  </div>
</div>
```

Using this annotation scheme the user can find the exact position of the needed mathematical expression in the web page.

TABLE1- The Added RDFa Annotations to the Presentation MathML.

| RDFa Field | Meaning |
|---|---|
| Anchor link | A link to the equation in the web page (e.g., www.example.com/lesson/lesson3.htm#equation2) |
| Value | The equation encoding in Content MathML |
| Source | The activity in e-learning in which the equation is found (e.g., Lesson) |
| Category | The mathematical equation category according to OpenMath (e.g., cosine is a trigonometric function) |
| Description | The uses of the mathematical equation (e.g., distance between 2 points) |

The following is a description of different fields in RDFa annotation (see Table1). The anchor link points where is the mathematical equation located in the web page. This link is composed of the URL address of the e-learning web page along with anchor ID for the mathematical equation in the web page. The anchor ID is a serial number automatically generated for each annotated mathematical equation in the current web page. The link can be used for a direct access to the location of the mathematical equation in the web page.

The source of the mathematical equation is extracted from the URL address of the current component where the mathematical equation was added (e.g., lesson, quiz, wiki). This information can be used later on to perform a more directed search by searching only in specific sections of the e-learning contents (e.g., search for quadratic equation in lessons only).

The value represents the content MathML encoding of the mathematical equation. Kohlhase and Sucan (Kohlhase & Sucan, 2006) mentioned that usually math web search uses content MathML as its basis. This encoding is obtained using a converter which converts from presentation MathML into contents MathML (for details see (Doush, Alkhateeb, & Maghayreh, 2010)). The use of content MathML can help in achieving a more structured unambiguous searching of the mathematical contents.

The category field represents the equation classification according to the OpenMath content dictionary ontology. The OpenMath ontology can be used to infer the mathematical equation category according to the classes and relations used in the Content Dictionary ontology. The reasoning can be performed according to the structure of the content MathML of the mathematical equation.

The use of mathematical equation is saved in the description field. Why scientists and mathematicians use the mathematical equation is entered by the mathematical contents creator according to the common uses of the equation in the field of study. The description can be helpful in cases where the user don't know the structure of the mathematical equation but s/he knows what the equation can be used for.

*G. Indexing of Mathematical Equations*

Indexing is the process of collecting, analyzing, and storing data to facilitate fast and accurate information retrieval (finding relevant documents for a search query). Without an index, the search engine has to scan all of the documents in the corpus, which would require a considerable amount of time (Baeza-yates & Ribeiro-Neto, 1999).

The indexing of the mathematical equations (see Figure3) needs to allow the user to perform different types of searching:

- When the user is searching for a mathematical expression s/he usually are looking for equations with similar structure and do not care about the variables used in the mathematical equation. This means that the retrieval system needs to return the mathematical equation in the search results even if it does not contain any of the variable names used in the search. For example, assume that the user query was "X = Y + Z - W" then the search should be performed on the equation "term = term + term - term".

- In some cases the user needs to retrieve the mathematical equation with exact term or variable names. In this case we have to give the user the option to search for the mathematical equation with exact variable names.

- Sometimes the user does not know the structure of the mathematical equation, but s/he knows what is the common name of the equation. Searching for a common name of the mathematical equation (e.g., Newton equation) needs to retrieve the mathematical equations with a structure similar to a defined equation with that common name.

- In some cases the user is interested in searching for equations under a large category (e.g., polynomial equations). The classification of equations into these categories relies on a mathematical ontology called OpenMath Content Dictionaries.

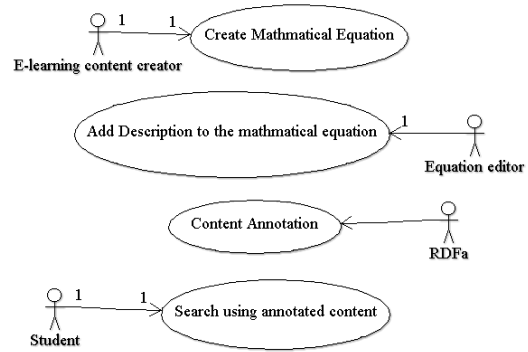

Figure4- A Use Case Diagram for the Indexing of Mathematical Equation.

The indexing of mathematical equations need to be performed by giving weights to the different operations applied to the equation terms (e.g., +, -). Order of evaluation is used to evaluate the weighting of terms in the mathematical equation.

The order of the operation in the equation affects the weight value used for the indexing as the order of the operations is an important criterion used when searching for the mathematical equations. The weight represents the value to be searched for when we are looking for a mathematical equation with similar structure and order of operations.

Another weight for the same equation can be calculated by including the variable names in the weight calculations. This weight value can be used for exact equation matching between the query and the indexed equations.

The user should have the ability to search for mathematical equations with the exact terms. The search needs to identify similarity distance between the searched equation and the indexed equations. The number of matched operations with correct order can be used as a weight to perform a partial matching for the mathematical operation.

In our context the documents represent a collection of mathematical expressions, and the query represents the user need to retrieve one or more expressions with certain characteristics of interest.

In textual documents, words and phrases can be used as index terms, but for mathematical expressions, what will be the index terms? From the representation of mathematical expressions (described in earlier sections) we can see that the index terms can involve the operators in the expressions, variable names, and other words used to describe the meaning or the uses of the mathematical expression.

The Vector Space Model (VSM) of information retrieval (Baeza-yates & Ribeiro-Neto, 1999) can be used in our context. In this case, both of the mathematical expression and the query are represented as vectors where each vector consists of a set of index terms' weights. The weight of an index term in the vector space model depends on its frequency in the collection of the documents (for more details see (Baeza-yates & Ribeiro-Neto, 1999).

To decide whether a given mathematical expression is relevant to a given query or not we have to measure the similarity between the query and the expression. In the VSM the similarity is represented by the cosine of the angle between the two vectors representing the expression and the query. The system must return as a response to a query a list of expressions ordered according to their similarity with the query, with the highest comes first. This is similar to what is done in traditional information retrieval systems.

The question to be addressed is how the user will enter his query? A simple solution is to use a simple textbox and allow the user to use a flag followed by a set of index terms that can be used in expressions.

For example the user can enter the following query:

Math: x y z + *

This query indicates that the user is looking for an expression that contains some or all of the above terms.

The only thing we need here during indexing is a procedure that scans the representation of the expression and tokenizes it where each token can be a variable name or an operator. These tokens along with any other words on the description of the expression represent the set of the index terms. The weight of these terms can be evaluated using the same method used in the VSM.

The user can enter the query directly as described above or we can allow him to use an equation editor to enter the query and then a procedure is used to tokenize it as it was indicated earlier. Then the weight of the terms in the query can be evaluated using the same method used in the VSM. Once we have the two vectors it is easy to calculate the similarity to determine whether the given expression is relevant to the query or not.

In this method the following query is identical to the one shown in the above example

Math: x + y * z

This is because the method does not consider the order of operator evaluation in the expression. If we want to consider this case then there are two possible solutions. The first one will require some changes in the way of evaluating the weight for the index terms. This is due to the fact that the current version of the VSM considers the frequency of the index terms in any document not its order. Using this solution the index terms with the same order of the query will have larger similarity than other index terms.

The second solution is simpler and described next. The index terms can be either variable names or pairs where each pair consists of an operator and a number representing its order of evaluation.

For example, given the following expression

x + y * z

The index terms will be x, y, z, (+,2), (*,1)

Now how the user will enter the query? There are two choices similar to the ones highlighted above. The first choice is to use a textbox and ask the user to enter the operators in his query as pairs.

For example if the user enters the following query

Math: (+,1) (*,2)

Then we understand that he is looking for an expression where the first operator evaluated is + and/or the second operator is *. In this case the similarity between this query and the above expression is zero.

The above explained method may not be easy to use and it may lead to many mistakes, as it is hard for some users to write their queries in the paired-format according to the order of the evaluation. Another solution for entering the query is to allow the user to write the query in it is normal format using a text or equation editor and have the parser generate the paired-format.

*H. Searching for Mathematical Equation*

Using the proposed searching scheme the users can apply more semantic search. For example the user can search for algebraic equations in quizzes only. Also, by applying the semantic queries the system can answer specific user's queries, for example the user can search for the lessons with Newton's equations.

Searching in the system can be performed in two modes:

- Searching using the search box by turning on the math search for the equations (e.g., Math: polynomial equations, and then the user specifies using a check box that the search is on the lessons).

- Searching by using the mathematical equation editor to enter the mathematical expression in the query. This query is converted into weighted query which is then searched for in the index. The user can specify if s/he is interested in partial matching of the query, and also if s/he is interested in exact matching with the terms and variable names of the equation.

SPARQL (Prud'hommeaux & Seaborne, 2008) is a W3C recommendation language developed in order to query RDF data. A simple SPARQL query is expressed using a form resembling the SQL SELECT query:

SELECT B FROM U WHERE P

Where U is the URL of an RDF graph G to be queried, P is a SPARQL graph pattern (i.e., a pattern constructed over RDF graphs with variables) and B is a tuple of variables appearing in P. Intuitively, an answer to a SPARQL query is an instantiation of the variables of B by the terms of the RDF graph G such that the substitution of the values to the variables of P yields to a subset of the graph G.

The following SPARQL query modeling this information:

SELECT *

```
FROM  <RDF data>
WHERE  {
        ?resource hasLink ?Link .
        ?resource hasValue ?Value .
        ?resource hasSource ?Source .
        ?resource hasCategory ?Category .
        ?resource hasDescription NewtonEquation .
      }
```

Could be used to search for annotations of a resource containing Newton's equation. The link (anchor link) of the resource will be returned to allow the user directly to access the page that contains it.

It should be noticed that the resource description and the value are determined from the user query. More precisely, the searching process is achieved in two phases. In the first phase, the user enters keywords to be searched. These keywords are used to search the description and the value of the resource (e.g., equation) to be retrieved. While in the second phase, the value and the description of the resource are used in a SPARQL query to retrieve the annotations of the resource.

The information obtained in the second phase is extracted from RDF document that matches the query. The RDF document represents the fields (as shown in Table1) retrieved using RDFa extractors.

## IV. SYSTEM DESIGN

The first prototype of the system has been realized by including our ontology and annotation components (accessible OATS) into Moodle. In order to make the annotation process possible, our tool is included inside Moodle in a place visible to all web pages of the course. An accessible modified version of the Open Annotation and Tagging System (OATS) (Bateman, Farzan, Brusilovsky, & McCalla, 2006) has been developed.

The new OATS has been made accessible, by replacing all user interactions with keyboard controls and aural messages to perform the highlighting and tagging. This has been applied in order to make the system accessible for people with disabilities. The e-learning upper-ontology is opened by accessible OATS and is available to the user for the annotation process.

The e-learning ontology provides the concepts taxonomy needed to classify the information (e.g., math → algebra → addition quiz) in the e-learning system. The relations defined in the ontology can be used to infer new information about the learning materials. For example, the has_answer relation between a quiz and a lesson can be used to imply a link between the instances of these classes.

The system structure is depicted in Figure5. The implementation relies on several software components:

- Connector between Apache and Tomcat: to integrate Apache and Tomcat mod-proxy-ajp, a connector to connect Tomcat and Apache Httpd, is used.

- REST web services (Servlets and AJAX): a REST architecture is used to update the tagging and annotation system database.

- JENA API: this semantic web API is used to maintain the e-learning ontology.

- WAI ARIA: the different classes for Moodle ontology are displayed to the students using an accessible tree structure built using WAI ARIA.

We propose also adding an open source equation editor (e.g., MathCast ("MathCast, an open source equation editor. http://mathcast.sourceforge.net/home.html," 2010)) to Moodle. The equation editor is used to add mathematical equations to the e-learning contents with presentation MathML encoding.

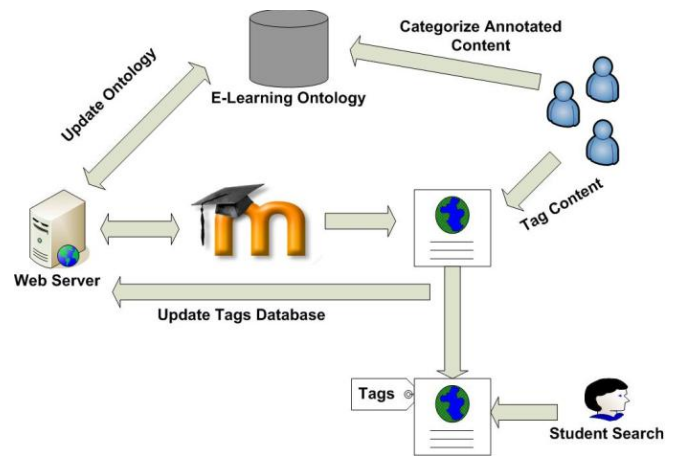

Figure5- The System Architecture.

## V. CONCLUSION AND FUTURE WORK

The proposed system uses collaborative tagging performed by the students in an online course to help other students. It relates different course components to each other, clarifying course materials by offering additional explanations, and providing connections between different course components. The context of the highlighted text can be used to select the ontology class for classifying the highlighted contents.

We presented investigation aimed at improving the process of searching for mathematical contents within e-learning systems. The goal of the proposed solution is to identify exactly where the mathematical expression is located in the web page by using RDFa annotations.

Using the proposed system the presentation MathML in the e-learning contents are automatically annotated with RDFa annotations using a pre-defined vocabulary to embed the semantics of the mathematical expression. The user query is then matched with the extracted RDFa annotations and the user is pointed to the list of URLs that have the mathematical formula.

The future work will include a comparison between the search using our proposed mathematical encoding and the regular text search. A user study for system evaluation is currently prepared to test the search enhancement when using the semantic search in e-learning. Another future direction is testing the system usability.